\documentclass[11pt]{amsart}
\usepackage{latexsym,amssymb,amsmath,amscd,amscd,amsthm,amsxtra}
\usepackage[dvips]{graphicx}
\usepackage{amsfonts}
\usepackage{amssymb}

\newtheorem{theorem}{Theorem}[section]
\newtheorem{corollary}[theorem]{Corollary}
\newtheorem{lemma}[theorem]{Lemma}
\newtheorem{proposition}[theorem]{Proposition}

\theoremstyle{definition}

\theoremstyle{remark}
\newtheorem*{remark}{Remark}
\newtheorem*{example}{Example}

\newcommand{\bH}{\mathcal{H}}

\newcommand{\NN}{\mathbb{N}}
\newcommand{\RR}{\mathbb{R}}
\newcommand{\ZZ}{\mathbb{Z}}
\newcommand{\AC}{\mathbb{C}}

\newcommand{\D}{\displaystyle}

\newcommand{\pf}{\noindent{\bf Proof.}\ }

\begin{document}

\title
{The Level Densities of Random Matrix Unitary Ensembles and their
Perturbation Invariability}

\author{Wang Zhengdong$^1$}
\address{1. School of mathematical science, Peking University,
Beijing, 100871, P. R. China} \email{zdwang@pku.edu.cn}

\author{Yan Kuihua$^{1,\,2}$}
\address{2. School of mathematics and
physics, Zhejiang Normal University, Jinhua city, Zhejiang
Province, 321004, P. R. China} \email{yankh@zjnu.cn}

\maketitle

\begin{abstract}

Using operator methods, we generally present the level densities
for kinds of random matrix unitary ensembles in weak sense. As a
corollary, the limit spectral distributions of random matrices
from Gaussian, Laguerre and Jacobi unitary ensembles are
recovered. At the same time, we study the perturbation
invariability of the level densities of random matrix unitary
ensembles. After the weight function associated with the 1-level
correlation function is appended a polynomial multiplicative
factor, the level density is invariant in the weak sense.

\end{abstract}

\vskip 0.5cm
\section{Introduction}
\vskip 0.5cm

\vskip 0.3cm In classical quantum mechanics, the statistical
properties of energy levels can be described by the $k$-level
correlation functions defined as (see Mehta\cite{Me})
$$
R_{_n\beta}^k(x_{_1},\,x_{_2},\,\cdots,\,x_{_k})=
\frac{n!}{(n-k)!}\int\cdots\int
P_{_{n\beta}}(x_{_1},\,x_{_2},\,\cdots,
\,x_{_n})\,dx_{_{k+1}}\cdots dx_{_n},
$$
where $P_{_{n\beta}}(x_{_1},\,x_{_2},\,\cdots,
\,x_{_n})=c_{_{n\beta}}\cdot \exp(-\beta H)$ is the joint
probability density function of $n$ eigenvalues
$x_{_1},\,x_{_2},\,\cdots, \,x_{_n}$ of a $n\times n$ random
matrix, $c_{_{n\beta}}$ is the normalized constant. $H$ is the
Hamiltonian of the logarithmical interacting $n$ particles system
on a straight line, which is given by constraining one-body
potential and logarithmic repulsive two-body potential, i.e.
$$
H=\sum_{i=1}^n V(x_{_i})-\sum_{1\leq i<j\leq
n}\log|x_{_i}-x_{_j}|, \quad x_{_i}\in \RR.
$$
In general, $V(x)$ is called potential function and $\beta$
Dyson's index. $\beta=1,\,2$ and $4$ are corresponding to the
orthogonal, unitary and symplectic ensembles respectively. When
$k=1$, $R_{_n\beta}^1(x)$ can be explained as the distribution
density of energy levels which can be found near by $x$. The level
density denoted by $\sigma_{_\beta}(x)$, which is a global
quantity, is defined by the limit of the $1$-level correlation
function $R_{_n\beta}^1(x)$. Then how to determine the level
density? It can be traced back to Wigner's pioneering
work\cite{Wig1,Wig2}. The results of early work are reviewed in
\cite{Me,Por}. Recently, there are many authors to concentrate on
this problem (See Spohn\cite{Sp}, Bai and Yin\cite{BY}, Nagao and
Wadati\cite{NW}, Haagerup and Thorbj$\o$rnsen\cite{HT},
Girko\cite{Gir}, Kiessling and Spohn\cite{KS},
Due$\tilde{n}$ez\cite{Due}, Ledoux\cite{Led}, etc.).

 \vskip 0.3cm Notice that given different or special
one-body potential $V(x)$, it will exhibited kinds of images for
us. As a matter of fact, in the case of classical Gaussian
ensembles, $\D V(x)=\frac{x^2}{2},\; x\in\RR$. The level density
is the famous ``semicircle law" first derived by
Wigner\cite{Wig1,Wig2}, i.e.
$$ \sigma_{_\beta}(x)=\left\{
\begin{array}{ll}
\frac{1}{\pi}\sqrt{2n-x^2} &x^2\leq 2n\\
0 &x^2\geq 2n.
\end{array}
\right.
$$

In addition, in the case of Laguerre ensembles, the level density
can be evaluated by a physical argument (See Bronk\cite{Bro}),
i.e.
$$ \sigma_{_\beta}(x)=\left\{
\begin{array}{ll}
\frac{1}{\pi\sqrt{x}}\sqrt{2n-x} &0<x\leq 2n\\
0 &x\geq 2n.
\end{array}
\right.
$$

In the case of Jacobi ensembles, the level density can also be
evaluated by a physical argument (See Leff\cite{Lef}), i.e.
$$ \sigma_{_\beta}(x)=\left\{
\begin{array}{ll}
\frac{n}{\pi\sqrt{1-x^2}} &-1<x<1\\
0 & otherwise.
\end{array}
\right.
$$

\vskip 0.3cm In the case of unitary ensembles, $R_{_{n2}}^1(x)$
can be expressed to a concise formula (see \cite{Me}, \cite{NW})
which is closely correlated to classical orthogonal polynomials,
i.e.
\begin{equation}\label{eq1}
R_{_{n2}}^1(x)=\sum_{m=0}^{n-1}p_{_m}^2(x)\varpi(x),
\end{equation}
where $\D \varpi(x)=\bar{c}\exp(-2V(x))$, $\bar{c}$ is the
normalized constant and $p_{_m}(x)$ be the $m$-order normalized
orthogonal polynomials associated with the normalized weight
function $\varpi(x)$, i.e. $\D \int p_{_m}(x)p_{_n}(x)\cdot
\varpi(x)\,dx=\delta_{_{mn}}$.

\vskip 0.3cm In \cite{HT}, Haagerup and Thorbj$\o$rnsen only
studied the Gaussian unitary ensemble(GUE) which was denoted by
$SGRM(n,\sigma^2)$ there. Using the following property of Hermite
polynomials
$$
H_{_k}(x+a)=\sum_{j=0}^kC_k^j(2a)^{k-j}H_{_j}(x),\quad a\in\RR,
$$
the authors directly obtained an equality for the complex Laplace
transform of $\D\frac{1}{n}R_{_{n2}}^{1}(x)$, i.e.
\begin{align*}
\int_{\RR}\exp(sx)\Bigl(\frac{1}{n}R_{_{n2}}^{1}(x) \Bigr)\,dx
&=\int_{\RR}\exp(sx)\Bigl(\frac{1}{n\sqrt{\pi}}\sum_{k=0}^{n-1}
\hat{H}_{_k}^2(x)e^{-x^2}\Bigr)\,dx\\
&=\exp(\frac{s^2}{4})\Phi(1-n,\,2;\,-\frac{s^2}{2}),\quad s\in
\AC,
\end{align*}
where $\hat{H}_{_k}(x)$ is the k-order normalized Hermite
polynomial, $\Phi(a,\,c;\,x)$ is the confluent hyper-geometric
function(CHGF) with parameters $a$ and $c$. Then they gave a short
proof of Wigner's semicircle law.

\vskip 0.3cm As we know, the CHGFs are complicated series
expansions. In \cite{Led}, Ledoux pushed forward the investigation
by Haagerup and Thorbj$\o$rnsen and only concentrated on the
differential aspects of CHGFs. The author constructed an abstract
framework of Markov diffusion generators, and in which derived the
basic differential equations on Laplace transforms of
$p_{_m}^2(x)\varpi(x)$.

\vskip 0.3cm Using the recurrence formula of Hermite polynomials
and the uniform integrability of random variable sequence, by the
obtained differential equation, the author showed that
$$
\lim_{m\rightarrow \infty}\int
f\Bigl(\frac{x}{2\sqrt{m}}\Bigr)\frac{1}{m}\sum_{k=0}^{m-1}p_{_k}^2(x)
\varpi(x)\,dx=E\Bigl(f(\sqrt{X}\,Y)\Bigr), \text{for all}\; f\in
C_{_b}(\RR)
$$
which determines the level density of GUE in the weak sense, where
$X$ and $Y$ are two independent random variables with the uniform
distribution on $[0,\,1]$ and the arcsine distribution on
$(-1,\,+1)$ respectively. By the analogous technique, the author
also obtained the level densities of Laguerre and Jacobi unitary
ensembles respectively.

\vskip 0.3cm In this paper, we will generally deal with the
problem for kinds of unitary ensembles (i.e. $\beta=2$) by
operator method. It is no confusion to omit the subscript 2 in the
below text. Moreover, as we will see, this method can effectively
be used to study the perturbation invariability of level density.

\vskip 0.3cm It is well known that the normalized orthogonal
polynomials $p_{_{n}}(x)$ satisfy the following recursion formula
(see section 2 for details)
$$
xp_{_{n}}(x)=\alpha_{_n}p_{_{n+1}}(x)+\beta_{_n}p_{_{n}}(x)
+\gamma_{_n}p_{_{n-1}}(x) ,\quad n=1,\,2,\,3,\,\cdots.
$$

\vskip 0.3cm In the paper, we assume that $\alpha_{_n}$ and
$\beta_{_n}$ satisfy the following {\bf exponential growth
conditions:}
\begin{equation}\label{eq2}
\alpha_{_n}=\xi n^t(1+\xi_{_n}), \quad \beta_{_n}=\zeta
n^t(1+\zeta_{_n})+~\eta_{_n},
\end{equation}
where $\xi\neq 0,\, \zeta\geq 0,\,0\leq t\leq 1$ are constants and
$\D\lim_{n\rightarrow \infty}\xi_{_n}=\lim_{n\rightarrow
\infty}\zeta_{_n}=\lim_{n\rightarrow \infty}\eta_{_n}=~0$.

\vskip 0.3cm Note that the classical Hermite, Laguerre and Jacobi
polynomials all satisfy exponential growth conditions (see section
2).

\vskip 0.3cm We define the ``ascending",``equilibrating" and
``descending" operators as follow,
$$
A_{_+}p_{_{n}}(x)=\alpha_{_n}p_{_{n+1}}(x),\quad
A_{_0}p_{_{n}}(x)=\beta_{_n}p_{_{n}}(x),\quad
A_{_-}p_{_{n}}(x)=\gamma_{_n}p_{_{n-1}}(x).
$$

In section 3, we use these operators to obtain the moment
inequality of the probability density $
\sigma_{_{n}}(x)=\frac{\sqrt{D_{_{n}}}}{n}
R_{_{n}}^{1}(x\sqrt{D_{_{n}}})$ and to show that the limit of k-th
moment $M_{_{n}}^{(k)}$ of $\sigma_{_{n}}(x)$ exists (see theorem
\ref{moment-limit1}), where $D_{_{n}}$ is the 2nd moment of
$\D\frac{1}{n} R_{_{n}}^{1}(x)$. Further more, by considering the
limit of characteristic functions of $\sigma_{_n}(x)$, we obtain
the main result in section 3, i.e. if there exists a probability
density $\sigma(x)$ with k-th moments $\D
M^{(k)}=\lim_{n\rightarrow\infty} M_{_n}^{(k)}$, then (see theorem
\ref{limti-density1})
$$\sigma_{_n}(x)\stackrel{w}{\longrightarrow}\sigma(x),\quad
n\rightarrow \infty,$$ where $w$ means weak convergence.

\vskip 0.3cm In section 4, the perturbation invariability of level
density is studied. We will consider another weight function
$\hat{\varpi}(x)$ by appending a polynomial multiplicative factor
$p^2(x)$ to $\varpi(x)$, which uniquely determines a family of
orthonormal polynomials $\{\hat{p}_{_m}(x)\}$. Accordingly,
$1$-level correlation function is $\D
\hat{R}_{_{n}}^{1}(x)=\sum_{j=0}^{n-1}\hat{p}_{_j}^2(x)p^2(x)\varpi(x).
$ We will show that the probability density
$\hat{\sigma}_{_{n}}(x)=\frac{\sqrt{D_{_{n}}}}{n}
\hat{R}_{_{n}}^{1}(x\sqrt{D_{_{n}}})$ is still weakly convergent
to $\sigma(x)$.

\vskip 0.5cm
\section{Orthogonal
Polynomials}

\vskip 0.3cm In this section, we briefly introduce the orthogonal
polynomials. We would like to lead the readers to refer
\cite{DX,NU,Sze}. Let us start by quoting a classical result as
follow.
\begin{theorem}\label{recurrence}
The following relation holds for the normalized orthogonal
polynomials $p_{_{n}}(x)$ with normalized weight function
$\varpi(x)$:
$$
xp_{_{n}}(x)=\alpha_{_n}p_{_{n+1}}(x)+\beta_{_n}p_{_{n}}(x)
+\gamma_{_n}p_{_{n-1}}(x) ,\quad n=1,\,2,\,3,\,\cdots
$$
where $\alpha_{_n},\,\beta_{_n}$ and $\gamma_{_n}$ are constants
which can be expressed in terms of the coefficients $a_{_n}$ and $
b_{_n}$ of the highest terms in $\D
p_{_n}(x)=a_{_n}x^n+b_{_n}x^{n-1}\cdots,\; a_{_n}\neq 0$, i.e.
$$
\alpha_{_n}=\frac{a_{_n}}{a_{_{n+1}}},\quad
\beta_{_n}=\frac{b_{_n}}{a_{_n}}-\frac{b_{_{n+1}}}{a_{_{n+1}}},\quad
n=0,\,1,\,\cdots,
$$
$$
\gamma_{_n}=\alpha_{_{n-1}},\quad \gamma_{_0}=0,\quad
n=1,\,2,\,\cdots.
$$
\end{theorem}

\vskip 0.3cm If we regard the multiplication by $x$ as an operator
$A_{_x}$, clearly the operator $A_{_x}$ plays
``ascending",``equilibrating" and ``descending" roles when it acts
on the $n$-order orthogonal polynomial $p_{_n}(x)$, i.e.
$$
A_{_x}=A_{_+}+A_{_0}+A_{_-},
$$
where $A_{_+},\,A_{_0}$ and
$A_{_-}$ are called ``ascending",``equilibrating" and
``descending" operators respectively and defined by
$$
A_{_+}p_{_{n}}(x)=\alpha_{_n}p_{_{n+1}}(x),\quad n=0,\,1,\,\cdots,
$$
\begin{equation}\label{eq3}
A_{_0}p_{_{n}}(x)=\beta_{_n}p_{_{n}}(x),\quad n=0,\,1,\,\cdots,
\end{equation}
$$
A_{_-}p_{_{n}}(x)=\gamma_{_n}p_{_{n-1}}(x),\quad
A_{_-}p_{_{0}}(x)=0,\quad n=1,\,2,\,\cdots.
$$

\vskip 0.3cm In the below text, we always assume that
$\alpha_{_n}$ and $\beta_{_n}$ satisfies the exponential growth
conditions (\ref{eq2}). Let us consider several classical
examples.

\vskip 0.3cm 1) Hermite polynomials

\vskip 0.2cm The weight function
$$
w(x)=\exp(-x^2), \quad x\in~\RR
$$
and
$$
\tilde{H}_{_n}(x)\equiv\frac{1}{\sqrt{2^nn!}}H_{_n}(x),
$$
where
$$
H_{_n}(x)=(-1)^n\exp(x^2)\frac{d^n}{dx^n}(\exp(-x^2))
$$
is the standard $n$-th Hermite polynomial which satisfies the
orthogonal relation
$$
\pi^{-\frac{1}{2}}\int_\RR H_{_k}(x)H_{_j}(x)\cdot
w(x)\,dx=2^kk!\delta_{_{kj}}.
$$

$\D \tilde{H}_{_n}(x)$ satisfies the following recurrence relation
$$
x\tilde{H}_{_n}(x)=\sqrt{\frac{n+1}{2}}\tilde{H}_{_{n+1}}(x)
+\sqrt{\frac{n}{2}}\tilde{H}_{_{n-1}}(x) ,\quad
n=1,\,2,\,3,\,\cdots
$$
$$
\tilde{H}_{_0}(x)=1,\quad \tilde{H}_{_1}(x)=\sqrt{2}\,x.
$$
Thus $\D\alpha_{_n}=\sqrt{\frac{n}{2}},\; \beta_{_n}=0$, i.e.
$\D\xi=\frac{1}{\sqrt{2}},\; \zeta=0,\; t=\frac{1}{2},\;
\xi_{_n}=\zeta_{_n}=\eta_{_n}=0$.

\vskip 0.3cm 2) Laguerre polynomials

\vskip 0.2cm The weight function
$$
w^{(a)}(x)=x^ae^{-x}, \quad x\in(0,\,\infty)
$$
and
$$
\tilde{L}_{_n}^{(a)}(x)\equiv\sqrt{\frac{\Gamma(k+1)}
{\Gamma(k+a+1)}}L_{_n}^{(a)}(x),
$$
where
$$
L_{_n}^{(a)}(x)=\frac{x^{-a}e^x}{n!}\frac{d^n}{dx^n}(x^{n+a}e^{-x})
$$
is the standard $n$-th Laguerre polynomial which satisfies the
orthogonal relation
$$
\int_0^\infty L_{_k}^{(a)}(x)L_{_j}^{(a)}(x)\cdot x^ae^{-x}\,dx
=\frac{\Gamma(k+a+1)}{\Gamma(k+1)}\delta_{_{kj}}.
$$

$\D \tilde{L}_{_n}^{(a)}(x)$ satisfies the following recurrence
relation
$$
x\tilde{L}_{_n}^{(a)}(x)=-\sqrt{(n+1)(n+1+a)}\tilde{L}_{_{n+1}}^{(a)}(x)
+(2n+a+1)\tilde{L}_{_n}^{(a)}(x)-\sqrt{n(n+a)}\tilde{L}_{_{n-1}}^{(a)}(x)
,
$$
$$
\tilde{L}_{_0}^{(a)}(x)=\frac{1}{\sqrt{\Gamma(a+1)}},\quad
\tilde{L}_{_1}^{(a)}(x)=\frac{-x+a+1}{\sqrt{\Gamma(a+2)}},\quad
n=1,\,2,\,3,\,\cdots.
$$
Thus $\D\alpha_{_n}=-\sqrt{(n+1)(n+1+a)},\; \beta_{_n}=2n+a+1$,
i.e. $\D\xi=-1,\; \zeta=~2,\; t=~1$,
$\D\xi_{_n}=\bigl(1+\frac{2+a}{n}+\frac{1+a}{n^2}\bigr)^{1/2}-1,\;
\zeta_{_n}=\frac{1+a}{2n},\; \eta_{_n}=0$.

\vskip 0.3cm 3) Jacobi polynomials

\vskip 0.2cm The weight function
$$
w^{(a,\,b)}(x)=(1-x)^a(1+x)^b, \quad x\in(-1,\,1)
$$
and
$$
\tilde{J}_{_n}^{(a,\,b)}(x)\equiv(c_{_n}^
{(a,\,b)})^{-\frac{1}{2}}J_{_n}^{(a,\,b)}(x),
$$
where
$$
c_{_k}^{(a,\,b)}=\frac{2^{a+b+1}}{2k+a+b+1}\frac{\Gamma(k+a+1)
\Gamma(k+b+1)} {\Gamma(k+1)\Gamma(k+a+b+1)}
$$
and
$$
J_{_n}^{(a,\,b)}(x)=\frac{1}{(1-x)^a(1+x)^b}\frac{(-1)^n}{2^nn!}
\frac{d^n}{dx^n}\bigl((1-x)^{n+a}(1+x)^{n+b}\bigr)
$$
is the standard $n$-th Jacobi polynomial which satisfies the
orthogonal relation
$$
\int_{-1}^1 J_{_k}^{(a,\,b)}(x)J_{_j}^{(a,\,b)}(x)\cdot
(1-x)^a(1+x)^b\,dx=c_{_k}^{(a,\,b)}\delta_{_{kj}}.
$$

$\D \tilde{J}_{_n}^{(a,\,b)}(x)$ satisfies the following
recurrence relation
\begin{align*}
x\tilde{J}_{_n}^{(a,\,b)}(x)=&\frac{2}{2n+a+b+2}
\sqrt{\frac{(n+1)(n+a+1)(n+b+1)(n+a+b+1)}{(2n+a+b+3)(2n+a+b+1)}}
\tilde{J}_{_{n+1}}^{(a,\,b)}(x)\\
&-\frac{a^2-b^2}{(2n+a+b)(2n+a+b+2)}
\tilde{J}_{_n}^{(a,\,b)}(x)\\
&+\frac{2}{2n+a+b}
\sqrt{\frac{n(n+a)(n+b)(n+a+b)}{(2n+a+b+1)(2n+a+b-1)}}
\tilde{J}_{_{n-1}}^{(a,\,b)}(x)
\end{align*}
Thus $\D\xi=\frac{1}{2},\; \zeta=0,\; t=0$. $\xi_{_n},\;
\zeta_{_n}$ and $\eta_{_n}$ have their respective expressions.

In particular, if $a=b=0$, then it is the Legendre polynomial. The
recurrence relation is
$$
x\tilde{J}_{_n}^{(0,\,0)}(x)=\frac{n+1}{\sqrt{(2n+3)(2n+1)}}
\tilde{J}_{_{n+1}}^{(0,\,0)}(x)+\frac{n}{\sqrt{(2n+1)(2n-1)}}
\tilde{J}_{_{n-1}}^{(0,\,0)}(x).
$$

%
%
%
%
%
%

\vskip 0.5cm
\section{The Level Densities}

\vskip 0.3cm \subsection{Main Results.}

\vskip 0.3cm In this section, we consider the limit of the global
$1$-level correlation functions $\D
R_{_n}^1(x)=\sum_{m=0}^{n-1}p_{_m}^2(x)\varpi(x)$ for unitary
ensembles, where $p_{_m}(x)$ is the m-th normalized orthogonal
polynomial associated with the normalized weight function
$\varpi(x)=\bar{c}\exp(-2V(x))$, $\bar{c}$ is the normalized
constant.

Specially, for the Gaussian unitary ensemble (GUE), $\D
V(x)=\frac{x^2}{2},\; x\in\RR$,
$$
\varpi(x)=\pi^{-\frac{1}{2}}w(x), \quad
p_{_m}(x)=\tilde{H}_{_m}(x), \quad x\in~\RR,
$$
where $\D \tilde{H}_{_m}(x)$ is the m-th normalized orthogonal
Hermite polynomial associated with the weight  function $\D
w(x)=\exp(-x^2)$.

\vskip 0.2cm For Laguerre unitary ensemble (LAUE), $\D
V(x)=\left\{
\begin{array}{ll}
-a\log x+x &0<x<\infty\\
\infty & otherwise
\end{array}
\right.$,
$$
\varpi(x)=\bar{c}w^{(2a)}(2x), \quad
p_{_m}(x)=\sqrt{2/\bar{c}}\tilde{L}_{_m}^{(2a)}(2x), \quad x\in
(0,\,\infty),
$$
where $\D \tilde{L}_{_m}^{(a)}(x)$ is the m-th normalized
orthogonal Laguerre polynomial associated with the weight function
$\D w^{(a)}(x)=x^ae^{-x}$.

\vskip 0.2cm For Jacobi unitary ensemble (JUE), $\D V(x)=\left\{
\begin{array}{ll}
-\log\bigl((1-x)^a(1+x)^b\bigr) &-1<x<1\\
\infty & otherwise
\end{array}
\right.$,
$$
\varpi(x)=\bar{c}w^{(2a,\,2b)}(x),\quad
p_{_m}(x)=\sqrt{1/\bar{c}}\tilde{J}_{_m}^{(2a,\,2b)}(x), \quad
x\in (-1,\,1),
$$
where $\D \tilde{J}_{_m}^{(a,\,b)}(x)$ is the m-th normalized
orthogonal Hermite polynomial associated with the weight  function
$\D w^{(a,\,b)}(x)=(1-x)^a(1+x)^b$. If $a=b=0$, it is the Legendre
unitary ensemble (LEUE).


\vskip 0.5cm Now it is no less of generality to consider the whole
real axis $\RR$. Denoted by $D_{_{n}}$ the 2nd moment of the
probability density $\D\frac{1}{n}R_{_{n}}^{1}(x)$. Then
\begin{align*}
D_{_{n}}&=\frac{1}{n}\int_\RR x^2 R_{_n}^1(x)\,dx
=\frac{1}{n}\sum_{j=0}^{n-1}\langle xp_{_j},\,xp_{_j}\rangle \\
&=\frac{1}{n}\sum_{j=0}^{n-1}(\alpha_{_j}^2+\beta_{_j}^2
+\gamma_{_j}^2).
\end{align*}
But $\gamma_{_j}=\alpha_{_{j-1}}$, hence
\begin{equation}\label{eq4}
D_{_{n}}=\frac{1}{n}\sum_{j=0}^{n-2}(2\alpha_{_j}^2+\beta_{_j}^2)+
\frac{1}{n}(\alpha_{_{n-1}}^2+\beta_{_{n-1}}^2).
\end{equation}

\vskip 0.3cm Let
$$
\sigma_{_{n}}(x)=\frac{\sqrt{D_{_{n}}}}{n}
R_{_{n}}^{1}(x\sqrt{D_{_{n}}}).
$$

Then the k-th moment $M_{_{n}}^{(k)}$ of probability density
$\sigma_{_{n}}(x)$ is
\begin{align*}
&M_{_{n}}^{(k)}=\int_\RR x^k\sigma_{_{n}}(x)\,dx
=\frac{1}{n\cdot(D_{_{n}})^{k/2}}\sum_{j=0}^{n-1}\int_{\RR}x^k
p_{_j}^2(x)\varpi(x)\,dx\\
&=\frac{1}{n\cdot(D_{_{n}})^{k/2}}\sum_{j=0}^{n-1}
\Bigl\langle x^kp_{_j},\,p_{_j}\Bigr\rangle_{_{L^2(\varpi)}}\\
&=\frac{1}{n\cdot(D_{_{n}})^{k/2}}\sum_{j=0}^{n-1}
\Bigl\langle (A_{_+}+A_{_0}+A_{_-})^kp_{_j},\,p_{_j}\Bigr\rangle_{_{L^2(\varpi)}}.\\
\end{align*}

Let $\Lambda_{_k}^i$ be an operator set composed of those terms in
the expansion of $(A_{_+}+A_{_0}+A_{_-})^k$, in which the
operators $A_{_+}$ and $A_{_-}$ exactly appear $i$ times. And note
that for all $\D T\not\in \bigcup_i\Lambda_{_k}^i$, $\D \langle
Tp_{_j},\,p_{_j}\rangle_{_{L^2(\varpi)}}=0$, thus we obtain the
following lemma

\vskip 0.3cm
\begin{lemma}\label{moment-equ1}
\begin{equation}\label{eq5}
M_{_{n}}^{(k)}=\frac{1}{n\cdot(D_{_{n}})^{k/2}}\sum_{j=0}^{n-1}
\sum_{i=0}^{[\frac{k}{2}]}\sum_{T\in \Lambda_{_k}^i} \Bigl\langle
Tp_{_j},\,p_{_j}\Bigr\rangle_{_{L^2(\varpi)}}.
\end{equation}
\end{lemma}

\vskip 0.3cm Moreover, set $\D
r_{_1}(k)=\sum_{i=0}^{[\frac{k}{2}]}C_k^iC_{k-i}^i\xi^{2i}
\zeta^{k-2i},\, r_{_2}(k)=(2\xi^2+\zeta^2)^{k/2}$ and $$
M^{(k)}=\frac{r_{_1}(k)}{r_{_2}(k)}\cdot\frac{(2t+1)^{k/2}}{(kt+1)},$$
then we have the following limit theorem for the k-th moment
$M_{_{n}}^{(k)}$.

\begin{theorem}\label{moment-limit1}
Under the exponential growth conditions,
\begin{equation}\label{eq6}
\lim_{n\rightarrow \infty}M_{_{n}}^{(k)}=M^{(k)},\quad \text{for
any}\; k\in \ZZ^+.
\end{equation}
\end{theorem}

The proof is based on elaborate estimations of $M_{_{n}}^{(k)}$.
Here we firstly consider the classical cases instead of being
anxious to verify this theorem.

\begin{example}
\vskip 0.3cm For GUE, $\D\xi=\frac{1}{\sqrt{2}},\; \zeta=0,\;
t=\frac{1}{2}$, so $\D
M^{(2m)}=\frac{C_{_{2m}}^m}{m+1},\;M^{(2m+1)}=0$. Moreover,
$M^{(k)}$ is exactly the k-th moment of density function
$$
\sigma_{_G}(x)=\left\{
\begin{array}{ll}
\frac{1}{2\pi}\sqrt{4-x^2} &x^2\leq 4\\
0 &x^2\geq 4.
\end{array}
\right.
$$

\vskip 0.3cm For LAUE, $\D\xi=-1,\; \zeta=~2,\; t=~1$, so $\D
M^{(k)}=\frac{C_{_{2k}}^k}{2^{k/2}(k+1)}$. Moreover, $M^{(k)}$ is
exactly the k-th moment of density function
$$
\sigma_{_L}(x)=\left\{
\begin{array}{ll}
\D\frac{\sqrt{2\sqrt{2}-x}}{\pi\sqrt{2x}} &0<x\leq 2\sqrt{2}\\
0 &x>2\sqrt{2}.
\end{array}
\right.
$$

\vskip 0.3cm For JUE, $\D\xi=\frac{1}{2},\; \zeta=0,\; t=0$, so
$\D M^{(2m)}=\frac{C_{_{2m}}^m}{2^{m}},\; M^{(2m+1)}=0$. Moreover,
$M^{(k)}$ is accordingly the k-th moment of density function
$$
\sigma_{_J}(x)=\frac{1}{\pi\sqrt{1-x^2}},\quad -1<x<1.
$$

\end{example}

\vskip 0.3cm Now suppose that there exists a probability density
$\sigma(x)$ with k-th moments $M^{(k)}$. Then we have

\vskip 0.3cm
\begin{theorem}\label{limti-density1} Under the exponential growth conditions,
\begin{equation}\label{eq8}
\sigma_{_n}(x)\stackrel{w}{\longrightarrow}\sigma(x),\quad
n\rightarrow \infty.
\end{equation}
where $w$ means weak convergence. In general, $\sigma(x)$ is
called a level density.
\end{theorem}

\vskip 0.3cm
\begin{corollary}\label{level-density} The level densities of
Gauss, Laguerre and Jacobi unitary ensembles are
$\sigma_{_G}(x),\,\sigma_{_L}(x)$ and $\sigma_{_J}(x)$
respectively in weak sense.
\end{corollary}

By the above example and theorem \ref{limti-density1}, it is
obvious.

\vskip 0.3cm
\subsection{Proof of main results}

Now we begin to verify the main results. First of all, let us
consider the 2nd moment $D_{_n}$ of
$\D\frac{1}{n}R_{_{n}}^{1}(x)$. If $\zeta>0$, set
\begin{equation}\label{defuj}
u_{_j}=\max_{j-k\leq m\leq
j+k}\Bigl\{|\xi_{_m}|,\,|\zeta_{_m}|+\frac{|\eta_{_m}|}{\zeta
m^t}\Bigr\},
\end{equation}
then $\D\lim_{j\rightarrow \infty}u_{_j}=0$. It is
no less of generality to assume $u_{_j}<1$. Thus by the
exponential growth conditions (\ref{eq2}) and equality
(\ref{eq4}),
\begin{equation}\label{eq9}
\frac{2\xi^2+\zeta^2}{n}\sum_{j=0}^{n-2}j^{2t}(1-u_{_j})^2 \leq
D_{_n}\leq
\frac{2\xi^2+\zeta^2}{n}\sum_{j=0}^{n-1}j^{2t}(1+u_{_j})^2.
\end{equation}
If $\zeta=0$, one can easily choose another proper infinitesimal
sequence $u_{_j}$ such that the above inequality is still valid.
By (\ref{eq9}), it is easy to see that
\begin{equation}\label{eqdn2}
\frac{D_{_n}}{n^{2t}}=O(1).
\end{equation}
\vskip 0.3cm {\it Proof of theorem \ref{moment-limit1}.} Firstly,
we come to verify a result about series limit. That is for any
sequence $x_{_n}$ and $y_{_n}$, if $\D\lim_{n\rightarrow
\infty}x_{_n}=0$ and $y_{_n}>0$, $\D \sum_{n=0}^\infty
y_{_n}=\infty$, then
\begin{equation}\label{eq16}
\lim_{n\rightarrow \infty}\frac{\D\sum_{j=0}^{n}y_{_j}x_{_j}}{\D
\sum_{j=0}^{n}y_{_j}}=0.
\end{equation}
Indeed, $\D\forall \varepsilon>0,\, \exists N\in\NN$, as $n\geq
N,\, |x_{_n}|<\frac{\varepsilon}{2}.$ Thus
$$
\Bigl|\frac{\D\sum_{j=0}^{n-1}y_{_j}x_{_j}}{\D
\sum_{j=0}^{n}y_{_j}}\Bigr|\leq\frac{\D
\sum_{j=0}^{N}y_{_j}|x_{_j}|}{\D
\sum_{j=0}^{n}y_{_j}}+\frac{\varepsilon}{2}\cdot\frac{\D
\sum_{j=N}^{n}y_{_j}}{\D
\sum_{j=0}^{n}y_{_j}}\leq\frac{\varepsilon}{2}+
\frac{\varepsilon}{2}=\varepsilon.
$$

Now let us set two cases to verify the theorem.

\vskip 0.3cm {\it Case 1.} $\zeta>0$.

\vskip 0.3cm  By exponential growth conditions (\ref{eq2}) and the
definition (\ref{defuj}) of $u_{_{j}}$, we see that
\begin{equation}\label{alphan1}
\max_{j-k\leq m\leq j+k}{\{|\alpha_{_m}|\}}\leq |\xi|
(j+k)^t(1+u_{_j}),\quad \max_{j-k\leq m\leq
j+k}{\{\beta_{_m}\}}\leq \zeta (j+k)^t(1+u_{_j}).
\end{equation}
Thus for $j>k$ and $\D T\in \Lambda_{_k}^i$,
\begin{align*}
\langle Tp_{_j},\,p_{_j}\rangle &\leq \Bigl(\max_{j-k\leq m\leq
j+k}{\{|\alpha_{_m}|\}}\Bigr)^{2i} \Bigl(\max_{j-k\leq m\leq
j+k}{\{\beta_{_m}\}}\Bigr)^{k-2i}\\
&\leq\Bigl(|\xi|(j+k)^t(1+u_{_{j}})\Bigr)^{2i}
\Bigl(\zeta(j+k)^t(1+u_{_{j}})\Bigr)^{k-2i}.
\end{align*}
And then
\begin{equation}\label{eq10}
\langle
Tp_{_j},\,p_{_j}\rangle\leq\xi^{2i}\zeta^{k-2i}(j+k)^{kt}(1+u_{_{j}})^k.
\end{equation}

\vskip 0.3cm Therefore by lemma \ref{moment-equ1} and inequalities
(\ref{eq10}),(\ref{eq9})-(\ref{eqdn2}),
\begin{equation}\label{eq11}
M_{_{n}}^{(k)}\leq \frac{r_{_1}(k)}{r_{_2}(k)}\cdot
\frac{\D\frac{1}{n}\sum_{j=0}^{n-1}(j+k)^{kt}(1+u_{_j})^k}{\D
\Bigl(\frac{1}{n}\sum_{j=0}^{n-2}j^{2t}(1-u_{_j})^2\Bigr)^{k/2}}
+O(\frac{1}{n^{kt+1}}),\quad \text{for}\; k\in\ZZ^+.
\end{equation}

 \vskip 0.3cm On the other hand, also by exponential growth
 conditions (\ref{eq2}) and the definition (\ref{defuj}) of $u_{_{j}}$,
\begin{equation}\label{alphan2}
\min_{j-k\leq m\leq j+k}{\{|\alpha_{_m}|\}}\geq |\xi|
(j-k)^t(1-u_{_j}) ,\quad \min_{j-k\leq m\leq
j+k}{\{\beta_{_m}\}}\geq\zeta (j-k)^t(1-u_{_j}).
\end{equation}
Then for $j>k$ and $\D T\in \Lambda_{_k}^i$,
\begin{align*}
\langle Tp_{_j},\,p_{_j}\rangle &\geq \Bigl(\min_{j-k\leq m\leq
j+k}{\{|\alpha_{_m}|\}}\Bigr)^{2i} \Bigl(\min_{j-k\leq m\leq
j+k}{\{\beta_{_m}\}}\Bigr)^{k-2i}\\
&\geq\Bigl(\xi(j-k)^t(1-u_{_{j}})\Bigr)^{2i}
\Bigl(\zeta(j-k)^t(1-u_{_{j}})\Bigr)^{k-2i}.
\end{align*}
Therefore,
\begin{equation}\label{eq12}
\langle
Tp_{_j},\,p_{_j}\rangle\geq\xi^{2i}\zeta^{k-2i}(j-k)^{kt}(1-u_{_{j}})^k.
\end{equation}

\vskip 0.3cm Thus by lemma \ref{moment-equ1} and inequalities
(\ref{eq9})-(\ref{eqdn2}) and (\ref{eq12}),
\begin{equation}\label{eq13}
M_{_{n}}^{(k)}\geq\frac{r_{_1}(k)}{r_{_2}(k)}\cdot
\frac{\D\frac{1}{n}\sum_{j=k}^{n-1}(j-k)^{kt}(1-u_{_j})^k}{\D
\Bigl(\frac{1}{n}\sum_{j=0}^{n-1}j^{2t}(1+u_{_j})^2\Bigr)^{k/2}}
+O(\frac{1}{n^{kt+1}}),\quad \text{for}\; n>k.
\end{equation}

Now let us come to consider the limit of the dexter series in
inequalities (\ref{eq11}) and (\ref{eq13}). By equality
(\ref{eq16}),
\begin{equation}\label{eq17}
\lim_{n\rightarrow
\infty}\frac{\D\frac{1}{n}\sum_{j=0}^{n-1}(j+k)^{kt}(1+u_{_j})^k}{\D
\Bigl(\frac{1}{n}\sum_{j=0}^{n-2}j^{2t}(1-u_{_j})^2\Bigr)^{k/2}}
=\lim_{n\rightarrow
\infty}\frac{\D\frac{1}{n}\sum_{j=0}^{n-1}(j+k)^{kt}}{\D
\Bigl(\frac{1}{n}\sum_{j=0}^{n-2}j^{2t}\Bigr)^{k/2}}
=\frac{(2t+1)^{k/2}}{(kt+1)}
\end{equation}
and
\begin{equation}\label{eq19}
\lim_{n\rightarrow
\infty}\frac{\D\frac{1}{n}\sum_{j=K_{_0}}^{n-1}(j-k)^{kt}(1-u_{_j})^k}{\D
\Bigl(\frac{1}{n}\sum_{j=0}^{n-1}j^{2t}(1+u_{_j})^2\Bigr)^{k/2}}
=\lim_{n\rightarrow
\infty}\frac{\D\frac{1}{n}\sum_{j=0}^{n-1}(j-k)^{kt}}{\D
\Bigl(\frac{1}{n}\sum_{j=0}^{n-2}j^{2t}\Bigr)^{k/2}}
=\frac{(2t+1)^{k/2}}{(kt+1)}.
\end{equation}
Then by inequalities (\ref{eq11}) and (\ref{eq13}) and equalities
(\ref{eq17})-(\ref{eq19}), we have
\begin{equation}\label{eq20}
\lim_{n\rightarrow \infty}M_{_{n}}^{(k)}
=\frac{r_{_1}(k)}{r_{_2}(k)}\cdot\frac{(2t+1)^{k/2}}{(kt+1)}=M^{(k)},\quad
\text{if}\;\, \zeta>0.
\end{equation}

\vskip 0.3cm {\it Case 2.} $\zeta=0$.

\vskip 0.3cm Set $\D v_{_j}=\max_{j-k\leq m\leq
j+k}\bigl\{|\xi_{_m}|,\,|\eta_{_m}|\bigr\}$. It is obvious that
$\D \lim_{j\rightarrow\infty}v_{_j}=0$. So it is no less of
generality to assume that $v_{_j}<1$. Thus by exponential growth
conditions ~(\ref{eq2}), for $0\leq j-k\leq m\leq j+k$,
\begin{equation}\label{alphabeta1}
|\xi|(j-k)^t(1-v_{_j})\leq |\alpha_{_m}|\leq
|\xi|(j+k)^t(1+v_{_j}),\quad |\beta_{_m}|\leq v_{_j}.
\end{equation}
Of course, $v_{_j}$ may be identity to zero. So we come to discuss
the estimation of $M_{_{n}}^{(k)}$ in terms of the parity of $k$.

\vskip 0.3cm (i) If $k$ is odd, then for $j>k$ and $\D T\in
\Lambda_{_k}^i$,
\begin{align*}
\Bigl|\langle Tp_{_j},\,p_{_j}\rangle\Bigr| &\leq\Bigl(
\xi(j+k)^t(1+v_{_j})\Bigr)^{2i}v_{_{j}}^{k-2i}\\
&\leq\xi^{2i}v_{_j}^{k-2i}(j+k)^{kt}(1+v_{_{j}})^k.
\end{align*}
Let $\D
\bar{v}_{_j}=\sum_{i=0}^{[\frac{k}{2}]}C_k^iC_{k-i}^i\xi^{2i}
v_{_j}^{k-2i}$, thus by inequality (\ref{eq9})-(\ref{eqdn2}) and
lemma~ \ref{moment-equ1},
\begin{equation}\label{eq14}
\Bigl|M_{_{n}}^{(k)}\Bigr|\leq \frac{1}{r_{_2}(k)}\cdot
\frac{\D\frac{1}{n}\sum_{j=0}^{n-1}(j+k)^{kt}(1+v_{_j})^k\bar{v}_{_j}}
{\D
\Bigl(\frac{1}{n}\sum_{j=0}^{n-2}j^{2t}(1-u_{_j})^2\Bigr)^{k/2}}
+O(\frac{1}{n^{kt+1}}),\quad \text{for}\; k\in\ZZ^+.
\end{equation}
Note that $\D \lim_{j\rightarrow\infty}\bar{v}_{_j}
=\lim_{j\rightarrow\infty}v_{_j}=0$. Then by equality (\ref{eq16})
and (\ref{eq17}),
$$
\lim_{n\rightarrow\infty}
\frac{\D\frac{1}{n}\sum_{j=0}^{n-1}(j+k)^{kt}(1+v_{_j})^k\bar{v}_{_j}}
{\D\Bigl(\frac{1}{n}\sum_{j=0}^{n-2}j^{2t}(1-u_{_j})^2\Bigr)^{k/2}}=0.
$$
Thus by inequalities (\ref{eq14}),
\begin{equation}\label{eq21}
\lim_{n\rightarrow \infty}M_{_{n}}^{(k)}=0=M^{(k)}\;
(\text{here}\;\, r_{_1}(k)=0).
\end{equation}

\vskip 0.3cm (ii) If k is even, then by inequality
(\ref{alphabeta1}), for $j>k$ and $\D T\in
\Lambda_{_k}^i,\;i\neq\frac{k}{2}$,
\begin{equation}\label{keventp1}
\Bigl|\langle Tp_{_j},\,p_{_j}\rangle\Bigr| \leq
\xi^{2i}v_{_j}^{k-2i}(j+k)^{kt}(1+v_{_{j}})^k
\end{equation}
and for $\D T\in \Lambda_{_k}^{k/2}$,
\begin{equation}\label{keventp2}
\xi^{k}(j-k)^{kt}(1-v_{_j})^k\leq\langle
Tp_{_j},\,p_{_j}\rangle\leq\xi^{k}(j+k)^{kt}(1+v_{_j})^k.
\end{equation}
Set $$
\tilde{M}_{_{n}}^{(k)}=\frac{1}{n\cdot(D_{_{n}})^{k/2}}\sum_{j=0}^{n-1}
\sum_{i\neq\frac{k}{2}}\sum_{T\in \Lambda_{_k}^i} \Bigl\langle
Tp_{_j},\,p_{_j}\Bigr\rangle$$ and $$
\bar{M}_{_{n}}^{(k)}=\frac{1}{n\cdot(D_{_{n}})^{k/2}}\sum_{j=0}^{n-1}\sum_{T\in
\Lambda_{_k}^{k/2}} \Bigl\langle Tp_{_j},\,p_{_j}\Bigr\rangle,$$
then by equality (\ref{eq5}),
\begin{align*}
M_{_{n}}^{(k)}=\tilde{M}_{_{n}}^{(k)}+\bar{M}_{_{n}}^{(k)}.
\end{align*}
It is completely analogous with the above discussion in the case
of odd k, we see that by inequalities (\ref{eq9}) and
(\ref{keventp1}) and equality (\ref{eq16}),
\begin{equation}\label{limtildem}
\lim_{n\rightarrow\infty}\tilde{M}_{_{n}}^{(k)}=0.
\end{equation}
But with respect to $\bar{M}_{_{n}}^{(k)}$, by inequalities
(\ref{eq9}) and (\ref{keventp2}), we have
\begin{equation}\label{estibarmsup}
\bar{M}_{_{n}}^{(k)}\leq \frac{C_k^{k/2}\xi^k}{r_{_2}(k)}\cdot
\frac{\D\frac{1}{n}\sum_{j=0}^{n-1}(j+k)^{kt}(1+v_{_j})^k}{\D
\Bigl(\frac{1}{n}\sum_{j=0}^{n-2}j^{2t}(1-u_{_j})^2\Bigr)^{k/2}}
+O(\frac{1}{n^{kt+1}})
\end{equation}
and
\begin{equation}\label{estibarmmin}
\bar{M}_{_{n}}^{(k)}\geq \frac{C_k^{k/2}\xi^k}{r_{_2}(k)}\cdot
\frac{\D\frac{1}{n}\sum_{j=k}^{n-1}(j-k)^{kt}(1-v_{_j})^k}{\D
\Bigl(\frac{1}{n}\sum_{j=0}^{n-2}j^{2t}(1+u_{_j})^2\Bigr)^{k/2}}
+O(\frac{1}{n^{kt+1}}).
\end{equation}
Then by equalities (\ref{eq17})-(\ref{eq19}) and inequalities
(\ref{estibarmsup})-(\ref{estibarmmin}),
\begin{equation}\label{limbarm}
\lim_{n\rightarrow\infty}\bar{M}_{_{n}}^{(k)}=
\frac{C_k^{k/2}\xi^k}{r_{_2}(k)}\cdot\frac{(2t+1)^{k/2}}{(kt+1)}.
\end{equation}
Therefore by equalities (\ref{limtildem}) and (\ref{limbarm}),
\begin{equation}\label{limmkeven}
\lim_{n\rightarrow\infty}M_{_{n}}^{(k)}=
\frac{C_k^{k/2}\xi^k}{r_{_2}(k)}\cdot\frac{(2t+1)^{k/2}}{(kt+1)}=
\frac{r_{_1}(k)}{r_{_2}(k)}\cdot\frac{(2t+1)^{k/2}}{(kt+1)}=M^{(k)}.
\end{equation}

\vskip 0.3cm Combining equalities (\ref{eq20}), (\ref{eq21}) and
(\ref{limmkeven}), we complete the proof of the theorem
\ref{moment-limit1}.

\vskip 0.3cm Next we come to verify the theorem
\ref{limti-density1}. In the first place, we have the following
rough estimation of k-th moment $M_{_{n}}^{(k)}$.

\begin{proposition}\label{rough-estimate}
Under the exponential growth conditions (\ref{eq2}), for any
$\varepsilon>0$ there exist an integer $K_{_0}(\varepsilon)$
independent on $k$ and $n$ such that
\begin{equation}\label{eq7}
\Bigl|M_{_{n}}^{(k)}\Bigr|\leq d_{_0}^k(\varepsilon k)^{kt},\quad
\text{for}\;k, n\geq K_{_0}(\varepsilon),
\end{equation}
where $d_{_0}$ is a constant independent on $k$, $n$ and
$\varepsilon$.
\end{proposition}

\vskip 0.3cm \pf By the exponential growth conditions (\ref{eq2}),
there is a constant $d_{_1}$ independent on $k$ and $n$, such that
$$
\quad |\alpha_{_n}|\leq d_{_1}n^t,\quad |\beta_{_n}|\leq
d_{_1}n^t.
$$
Then for any $\D T\in \Lambda_{_k}^i$,
\begin{equation}\label{eqtp}
\Bigl|\langle Tp_{_j},\,p_{_j}\rangle\Bigr|\leq
d_{_1}^{k}(j+k)^{kt}.
\end{equation}
And by inequality (\ref{eq9}), there is a constant $d_{_2}>0$
independent on $k$ and $n$, such that
\begin{equation}\label{eqdn}
D_{_n}\geq d_{_2}n^{2t}.
\end{equation}
Thus by lemma \ref{moment-equ1} and inequalities
(\ref{eqtp})-(\ref{eqdn}),
$$
\Bigl|M_{_{n}}^{(k)}\Bigr|\leq
\frac{\D\sum_{i=0}^{[\frac{k}{2}]}C_k^iC_{k-i}^i
}{\D(d_{_2})^{k/2}}\cdot
\frac{\D\sum_{j=0}^{n-1}d_{_1}^{k}(j+k)^{kt}}{\D n^{kt+1}}\leq
\frac{d_{_1}^{k}\D\sum_{i=0}^{[\frac{k}{2}]}C_k^iC_{k-i}^i
}{\D(d_{_2})^{k/2}}\cdot \Bigl(\frac{n+k}{n}\Bigr)^{kt}
$$
But $\D \sum_{i=0}^{[\frac{k}{2}]}C_k^iC_{k-i}^i\leq 3^k$. Set $\D
d_{_0}=\frac{3d_{_1}}{d_{_2}^{1/2}}$, then
$$
\Bigl|M_{_{n}}^{(k)}\Bigr|\leq
d_{_0}^k\Bigl(1+\frac{k}{n}\Bigr)^{kt}.
$$

\vskip 0.3cm Note that for all $\varepsilon>0$, set
$K_{_0}(\varepsilon)=[\frac{2}{\varepsilon}]+2$, then when
$k,n\geq K_{_0}(\varepsilon)$, $\D 1+\frac{k}{n}\leq
\frac{\varepsilon}{2}k+\frac{\varepsilon}{2}k\leq \varepsilon k$.
Consequently,
$$
\Bigl|M_{_{n}}^{(k)}\Bigr|\leq d_{_0}^k(\varepsilon k)^{kt}, \quad
\text{for}\;k,n\geq K_{_0}(\varepsilon).
$$
Thus we complete the proof of proposition~ \ref{rough-estimate}.

\vskip 0.3cm {\it Proof of theorem \ref{limti-density1}.} Let
$f_{_n}(\theta)$ and $f(\theta)$ be the characteristic functions
of $\sigma_{_n}(x)$ and $\sigma(x)$ respectively. It is sufficient
to show that
$$\lim_{n\rightarrow \infty}f_{_n}(\theta)=f(\theta).$$

Note that
\begin{align*}
\sum_{k=0}^{\infty}\frac{|\theta|^k}{k!}\int
|x|^k\sigma_{_n}(x)\,dx
&=\sum_{m=0}^{\infty}\frac{|\theta|^{2m}}{(2m)!} M_{_n}^{(2m)}+
\sum_{m=0}^{\infty}\frac{|\theta|^{2m+1}}{(2m+1)!}E(|X_{_n}|^{2m+1})\\
&\leq \sum_{m=0}^{\infty}\frac{|\theta|^{2m}}{(2m)!}M_{_n}^{(2m)}
+\sum_{m=0}^{\infty}\frac{|\theta|^{2m+1}}{(2m+1)!}
\Bigl(1+M_{_n}^{(2m+2)}\Bigr).
\end{align*}
Then by the rough estimation (\ref{eq7}) of $M_{_n}^{(k)}$ in
proposition \ref{rough-estimate}, for any $\theta$ we can choose
$\varepsilon>0$ such that
$$
\sum_{k=0}^{\infty}\frac{|\theta|^k}{k!}\int
|x|^k\sigma_{_n}(x)\,dx<\infty,\quad \text{for}\;n\geq
K_{_0}(\varepsilon).
$$
Thus we have
\begin{align*}
&f_{n}(\theta)=\int e^{i\theta
x}\sigma_{_n}(x)\,dx=\int\sum_{_{k=0}}^\infty\frac{(i\theta
x)^k}{k!}\sigma_{_n}(x)\,dx\\
&=\sum_{_{k=0}}^\infty\int\frac{(i\theta
x)^k}{k!}\sigma_{_n}(x)\,dx
=\sum_{_{k=0}}^\infty\frac{(i\theta)^{k}}{k!}M_{_n}^{(k)},\quad
\text{for}\;n\geq K_{_0}(\varepsilon).
\end{align*}
Analogously, it can be verified that $\D
f(\theta)=\sum_{_{k=0}}^\infty\frac{(i\theta)^{k}}{k!}M^{(k)} $.

Note that $0\leq t\leq 1$. So for any $\theta$ we can choose
$\varepsilon>0$ such that
$$\D\sum_{_{k=K_{_0}(\varepsilon)}}^\infty\frac{|\theta
d_{_0}|^k(\varepsilon k)^{kt}}{k!}<\infty.$$ Therefore by theorem
\ref{moment-limit1}, inequality (\ref{eq7}) and Lebesgue control
convergent theorem for series,
$$
\lim_{n\rightarrow
\infty}\sum_{_{k=0}}^\infty\frac{(i\theta)^{k}}{k!}M_{_n}^{(k)}=
\sum_{_{k=0}}^\infty\frac{(i\theta)^{k}}{k!}\lim_{n\rightarrow
\infty}M_{_n}^{(k)}=
\sum_{_{k=0}}^\infty\frac{(i\theta)^{k}}{k!}M^{(k)}.
$$
Thus we obtain
$$
\lim_{n\rightarrow \infty}f_{n}(\theta)=f(\theta).
$$
It completes the proof of theorem \ref{limti-density1}.

\vskip 0.5cm
\section{Perturbation Invariability}

\vskip 0.3cm Now Let $p(x)$ be a fixed l-order polynomial, denoted
by $\hat{p}_{_j}(x)$ the j-th normalized orthogonal polynomials
associated with the weight function
$\hat{\varpi}(x)=p^2(x)\varpi(x)$. By equality (\ref{eq1}), the
$1$-level correlation function is
$$
\hat{R}_{_{n}}^{1}(x)=\sum_{j=0}^{n-1}\hat{p}_{_j}^2(x)p^2(x)\varpi(x).
$$
Let
$$
\hat{\sigma}_{_{n}}(x)=\frac{\sqrt{D_{_{n}}}}{n}
\hat{R}_{_{n}}^{1}(x\sqrt{D_{_{n}}}).
$$

 Set
$$
\bH_{_n}=span\{p_{_0}(x),\,p_{_1}(x), \,\cdots,\, p_{_{n-1}}(x)\},
$$
then $\bH_{_n}$ is a subspace of $L^2(\RR,\,\varpi(x)dx)$ with $n$
dimensions. It is obvious that $\D
\{\hat{p}_{_0}(x)p(x),\,\cdots,\, \hat{p}_{_{n-l-1}}(x)p(x)\}$ is
a family of normalized orthogonal vectors in $\bH_{_n}$. We can
extend this set of vectors, such that it makes up of a normalized
orthogonal base of $\bH_{_n}$. Denoted it by
$$ \{e_{_0}^{(n)}(x),\,\cdots,\,
e_{_{l-1}}^{(n)}(x),\, \hat{p}_{_0}(x)p(x),\,\cdots,\,
\hat{p}_{_{n-l-1}}(x)p(x)\}.
$$

Let $P_{_n}$ be the projective operator from
$L^2(\RR,\,\varpi(x)dx)$ to $\bH_{_n}$. we construct an operator
$T_{_n}^{(k)}$ from $\bH_{_n}$ to itself as follows,
\begin{equation}\label{eq23}
T_{_n}^{(k)}=P_{_n}\circ A_{_x}^k,
\end{equation}
where $A_{_x}$
is multiplication by x.

\vskip 0.3cm Now we consider the $k$-moment $M_{_{n}}^{(k)}$ and
$\hat{M}_{_{n}}^{(k)}$ of probability density $\sigma_{_{n}}(x)$
and $\hat{\sigma}_{_{n}}(x)$ respectively.

\vskip 0.3cm
\begin{proposition}\label{momentequ1}
\begin{equation}\label{eq24}
 M_{_{n}}^{(k)}=\frac{Tr(T_{_n}^{(k)})}
{n\cdot(D_{_{n}})^{k/2}}.
\end{equation}
\end{proposition}

\pf
\begin{align*}
&M_{_{n}}^{(k)}=\int_\RR x^k\sigma_{_{n}}(x)\,dx
=\frac{1}{n\cdot(D_{_{n}})^{k/2}}\sum_{j=0}^{n-1} \Bigl\langle
x^kp_{_j},\,p_{_j}\Bigr\rangle_{_{L^2(\varpi)}}\\
&=\frac{1}{n\cdot(D_{_{n}})^{k/2}}\sum_{j=0}^{n-1} \Bigl\langle
A_{_x}^k(p_{_j}),\,P_{_n}(
p_{_j})\Bigr\rangle_{_{L^2(\varpi)}}\\
&=\frac{1}{n\cdot(D_{_{n}})^{k/2}}\sum_{j=0}^{n-1} \Bigl\langle
T_{_n}^{(k)}(p_{_j}),\, p_{_j}\Bigr\rangle_{_{L^2(\varpi)}}
=\frac{Tr(T_{_n}^{(k)})}{n\cdot(D_{_{n}})^{k/2}}.\\
\end{align*}

\vskip 0.3cm
\begin{lemma}\label{xf-esm}
\begin{equation}\label{eq25}
||A_{_x}f||_{_{L^2(\varpi)}}\leq 3N_{_n}||f||,\quad \text{for all
f}\in \bH_{_n},
\end{equation}
where $\D N_{_n}=\sup \limits_{0\leq i\leq
n-1}\{|\alpha_{_i}|,\,|\beta_{_i}|\}$.

\end{lemma}

\pf  Put $\D f=\sum_{i=0}^{n-1}c_{_i}p_{_i}$, then
\begin{align*}
||A_{_x}f||^2&=||\sum_{i=0}^{n-1}c_{_i}xp_{_i}||^2 =||\sum_ic_{_i}
(\alpha_{_i}p_{_{i+1}}+\beta_{_i}p_{_i}+\gamma_{_i}p_{_{i-1}})||^2\\
&=||\sum_i(
c_{_{i-1}}\alpha_{_{i-1}}+c_{_i}\beta_{_i}+c_{_{i+1}}\gamma_{_{i+1}})p_{_i}||^2\\
&=\sum_i|c_{_{i-1}}\alpha_{_{i-1}}+c_{_i}\beta_{_i}+c_{_{i+1}}\alpha_{_i}|^2\\
&\leq 3N_{_n}^2\sum_i(c_{_{i-1}}^2+c_{_i}^2+c_{_{i+1}}^2)\leq
9N_{_n}^2||f||^2.
\end{align*}
Therefore, $\D ||A_{_x}f||\leq 3N_{_n}||f||.$

\vskip 0.3cm
\begin{corollary}\label{multi-estimate}
\begin{equation}\label{eq26}
 ||A_{_x}^kf||\leq
3^k\Bigl(\prod_{j=0}^{k-1}N_{_{n+j}}\Bigr)||f||, \quad \text{for
all f}\in \bH_{_n}.
\end{equation}
\end{corollary}

\pf By the above proposition, it is obvious.

\vskip 0.3cm
\begin{proposition}\label{momentequ2}
\begin{equation}\label{eq27}
 \hat{M}_{_n}^{(k)}=M_{_{n}}^{(k)}+I,
\end{equation}
where
$$
I\equiv \frac{1}{n\cdot(D_{_{n}})^{k/2}}
\Bigl(\sum_{j=n-l}^{n-1}\Bigl\langle A_{_x}^k
(\hat{p}_{_j}p),\,\hat{p}_{_j}p\Bigr\rangle_{_{L^2(\varpi)}}
-\sum_{j=0}^{l-1}\Bigl\langle
T_{_n}^{(k)}(e_{_j}^{(n)}),\,e_{_j}^{(n)}
\Bigr\rangle_{_{L^2(\varpi)}}\Bigr).
$$
\end{proposition}

\pf
\begin{align*}
&\hat{M}_{_n}^{(k)}=\int_\RR x^k\hat{\sigma}_{_{n}}(x)\,dx
=\frac{1}{n\cdot(D_{_{n}})^{k/2}}\int_{\RR}x^k
\hat{R}_{_{n}}^{1}(x)\,dx\\
&=\frac{1}{n\cdot(D_{_{n}})^{k/2}}\sum_{j=0}^{n-1}\int_{\RR}x^k
\hat{p}_{_j}^2(x)p^2(x)\varpi(x)\,dx\\
&=\frac{1}{n\cdot(D_{_{n}})^{k/2}}\Bigl(\sum_{j=0}^{n-l-1}\int_{\RR}x^k
\hat{p}_{_j}^2(x)p^2(x)\varpi(x)\,dx+\sum_{j=0}^{l-1}\int_{\RR}x^k
(e_{_j}^{(n)}(x))^2\,dx\Bigr)\\
&\qquad+\frac{1}{n\cdot(D_{_{n}})^{k/2}}
\Bigl(\sum_{j=n-l}^{n-1}\int_{\RR}x^k
\hat{p}_{_j}^2(x)p^2(x)\varpi(x)\,dx-\sum_{j=0}^{l-1}\int_{\RR}x^k
(e_{_j}^{(n)}(x))^2\,dx\Bigr)\\
&=\frac{1}{n\cdot(D_{_{n}})^{k/2}}\Bigl(\sum_{j=0}^{n-l-1}
\Bigl\langle T_{_n}^{(k)}
(\hat{p}_{_j}p),\,\hat{p}_{_j}p\Bigr\rangle_{_{L^2(\varpi)}}
+\sum_{j=0}^{l-1} \Bigl\langle
T_{_n}^{(k)}(e_{_j}^{(n)}),\,e_{_j}^{(n)}\Bigr\rangle_{_{L^2(\varpi)}}
\Bigr)\\
&\qquad+\frac{1}{n\cdot(D_{_{n}})^{k/2}}
\Bigl(\sum_{j=n-l}^{n-1}\Bigl\langle A_{_x}^k
(\hat{p}_{_j}p),\,\hat{p}_{_j}p\Bigr\rangle_{_{L^2(\varpi)}}
-\sum_{j=0}^{l-1}\Bigl\langle
T_{_n}^{(k)}(e_{_j}^{(n)}),\,e_{_j}^{(n)}
\Bigr\rangle_{_{L^2(\varpi)}}\Bigr)\\
&=\frac{Tr(T_{_n}^{(k)})}{n\cdot(D_{_{n}})^{k/2}}+I\;.\\
\end{align*}
Then by proposition \ref{momentequ1}, we obtain the conclusion.

\vskip 0.5cm
\begin{theorem}\label{invariant-weight}
Under the exponential growth conditions,
\begin{equation}\label{eq28}
\lim_{n\rightarrow \infty}M_{_{n}}^{(k)}=\lim_{n\rightarrow
\infty}\hat{M}_{_n}^{(k)}, \quad \text{for any}\; k\in \ZZ^+.
\end{equation}
\end{theorem}

\vskip 0.3cm \pf By the exponential growth conditions and
corollary \ref{multi-estimate}, we see that there exist constants
$C_{_1},\,C_{_2}>0$ such that
\begin{equation}\label{eq29}
\Bigl|\Bigl\langle A_{_x}^k
(\hat{p}_{_j}p),\,\hat{p}_{_j}p\Bigr\rangle_{_{L^2(\varpi)}}\Bigr|\leq
||A_{_x}^k\hat{p}_{_j}p||\leq 3^kN_{_{n+l+k-2}}^k\leq C_{_1}n^{kt}
\end{equation}
and
\begin{equation}\label{eq30}
\Bigl|\Bigl\langle
T_{_x}^{(k)}(e_{_j}^{(n)}),\,e_{_j}^{(n)}\Bigr\rangle_{_{L^2(\varpi)}}\Bigr|\leq
||P_{_n}||\cdot||A_{_x}^ke_{_j}^{(n)}||\leq 3^kN_{_{n+k-1}}^k\leq
C_{_2}n^{kt}.
\end{equation}

\vskip 0.3cm Thus by proposition \ref{momentequ2}, as $n>>1$,
\begin{align*}
|I|\leq& \frac{1}{n\cdot(D_{_{n}})^{k/2}}\Bigl(\sum_{j=n-l}^{n-1}
C_{_1}n^{kt}+\sum_{j=0}^{l-1}C_{_2}n^{kt}\Bigr)\\
&=\frac{(C_{_1}+C_{_2})l}{n}
\cdot\frac{n^{kt}}{(D_{_{n}})^{k/2}}=O(\frac{1}{n}).
\end{align*}
Whereupon we obtain
$$\lim_{n\rightarrow \infty}M_{_{n}}^{(k)}=\lim_{n\rightarrow
\infty}\hat{M}_{_n}^{(k)}.$$

\begin{remark}
In the case of finite interval $(-s,\,s)\subset\RR$, note that the
operator $A_{_x}$ is bounded, so the inequalities (\ref{eq29}) and
(\ref{eq30}) in the above proof are naturally valid.
\end{remark}

\vskip 0.3cm Finally we have the following theorem which tells us
that after the weight function $\varpi(x)$ is appended a
polynomial multiplicative factor, the limit behavior of the
normalized 1-level correlation function is unaffected in the weak
sense.
\begin{theorem}\label{limti-density2} Under the exponential growth conditions,
$$\hat{\sigma}_{_n}(x)\stackrel{w}{\longrightarrow}\sigma(x),\quad
n\rightarrow \infty,$$ where $w$ means weak convergence.
\end{theorem}

\pf It is completely analogous to the proof of theorem
\ref{limti-density1}, here we omit it.


\end{document}